\documentclass[a4paper]{article}
\usepackage{amsmath}
\usepackage{amsfonts}
\usepackage{graphicx}
\usepackage{geometry}
\usepackage{pifont}

\begin{document}

\title{A Dynamic Programming Implemented $2 \times 2$  non-cooperative Game Theory Model for ESS Analysis}

 \author{Chen Shi \thanks{Department of Entomology and Program of Operations Research, the Pennsylvania State University, PA, 16801, USA}
  ,Fang Yuan \thanks{Department of Computer Sciences and Engineering and Program of Operations Research,the Pennsylvania State University, PA, 16801, USA}
   }

\date{}
 \maketitle

\begin{abstract}
Game Theory has been frequently applied in biological research
since 1970s. While the key idea of Game Theory is Nash
Equilibrium, it is critical to understand and figure out the
payoff matrix in order to calculate Nash Equilibrium. In this
paper we present a dynamic programming implemented method to
compute $2 \times2$ non-cooperative finite resource allocation
game's payoff matrix. We assume in one population there exists two
types of individuals, aggressive and non-aggressive and each
individual has equal and finite resource. The strength of
individual could be described by a function of resource
consumption in different development stages. Each individual
undergoes logistic growth hence we divide the development into
three stages: initialization, quasilinear growth and termination.
We first discuss the theoretical frame of how to dynamic
programming to calculate payoff matrix then give three numerical
examples representing three different types of aggressive
individuals and calculate the payoff matrix for each of them
respectively. Based on the numerical payoff matrix we further
investigate the evolutionary stable strategies (ESS) of the games.
\end{abstract}

{\bf Keywords:} Dynamic Programming, Finite Resource Allocation,
Game Theory,  Evolutionary Stable Strategy (ESS)

\section{Introduction}
Game Theory is originally introduced by John von Neumann and Oskar
Morgenstern (Neumann J and Morgenstern O. 1944) in 1944. Later
John F. Nash (Nash JF. 1950) has made significant contribution to
Game Theory by introducing Nash Equilibrium and proving every
finite game has mixed Nash Equilibrium, which becomes the core
idea of Game Theory. Since then Game Theory has been widely
applied in various disciplines such as social sciences (most
notably economics), political science, international relations,
computer science and philosophy. So far eight game theorists have
won Nobel prizes in economics. John M. Smith in 1973 formalized
another central concept in Game Theory called the Evolutionary
Stable Strategy (ESS, Smith JM. 1973). Various work has been done
using ESS to investigate the behavior and evolutionary path of
animals ( Takada T and Kigami J. 1991, Crowley PH. 2000, Nakamaru
M and Sasaki A. 2003, Matsumura S and Hayden TJ. 2005, Wolf N and
Mangel M. 2007, Krivan V, Cressman R and Schneider C. 2008,
Hamblin S and Hurd PL. 2009).
\\
The simplest yet most common case of ESS game is $2 \times 2$
non-cooperative game. Multiple player game is more realistic but
according to Poincare-Bendixson theorem, multiplayer dynamic
system would result in chaos hence we only consider $2 \times 2$
games (Yi T et al. 1997). While the key idea in Game Theory is
Nash Equilibrium, we must define the payoff matrix very carefully
in order to compute Nash Equilibrium. However, most of the
research regarding ESS has arbitrarily assigned payoff matrix. To
overcome this problem, we want to use some realistic quantities to
define payoff matrix for the game. M-Gibbons has shown a neighbor
intervention model (Mesterton-Gibbons M, Sherratt TN. 2008) and
Luther further discussed whether food is worth fighting for
(Luther RM, Broom M and Ruxton GD. 2007). Just has studied the
aggressive loser in an ESS game (Just W, Morris MR and Sun X.
2006). Inspired by their ideas, we will use food source to compute
the payoff of aggressive and non-aggressive players in a $2 \times
2$ game. \\
While we assume the food source is finite and equivalent to any
member in the population, it is natural to use dynamic programming
(DP) to figure out the optimal foraging strategy as a resource
allocation problem. Animal growth rate is a logistic curve and we
divide the whole growth process into three distinct stages:
initialization, quasilinear growth and termination. In different
stages, the payoff is a linear function of food source with
different slope and our goal is to determine the maximum total
payoff at the end of growth using dynamic
programming. \\
Al-Tamimi has suggested using dynamic programming to implement
Game Theory model for designing (Al-Tamimi A, Abu-Khalaf M and
Lewis FL. 2007) but their model is zero-sum. We will first present
a more realistic general sum game framework, then discuss three
different types of aggressive players, calculate the numerical
payoff matrix for each case and determine the ESS for them. Our
work is the first of this kind to combine dynamic programming and
Game Theory, two different optimization tools together to solve
real biological problem.

\section{Defining the Model}
A typical $2 \times 2$ non-cooperative general sum game has the
following form where $P_{ji}$ defines the payoff of player \emph{\it{i}} in \emph{\it{j}}th strategy combination.\\
\begin{center}
\begin{tabular}{c c c}
  \hline
  Strategy & Non-aggressive & Aggressive \\
  \hline
  Non-aggressive & $(P_{111}, P_{112})$ & $(P_{121}, P_{122})$ \\
  Aggressive & $(P_{211}, P_{212})$ & $(P_{221}, P_{222})$ \\
  \hline
\end{tabular}
\\
Table 1. Payoff Matrix of non-cooperative general sum game
\end{center}
$P_{ijk}$ denotes the payoff of player  \textit{k} when it uses
strategy \textit{i} and its components uses  \textit{j}. Here we
have two types of strategies: aggressive (2) and non-aggressive
(1). Aggressive players would fight their neighbor and try to get
their resources. Non-aggressive players only concentrate on their
own food source and never fight back even when they are attacked.
However, if two aggressive players meet, it would result in a
severe fight and both players are terribly hurt. This definition
is similar to that of "Chicken-Dare" or "Hawk-Dove" game. The Nash
Equilibrium is defined as:\\
$\textbf{Definition 1. } x \in \Theta$ is a Nash Equilibrium if $x
\in \tilde{\beta} (x)$, where $\Theta$ is the mixed strategy space
and $\tilde {\beta}$ is
the mixed strategy best response correspondence.\\
Because this is a $2 \times 2$ finite symmetric game,
$\Delta^{NE}\neq \emptyset$ by Kakutani's Theorem. Next we switch
to a population perspective and define Evolutionary Stable
Strategy (ESS) as follows:\\
$\textbf{Definition 2. } x \in \Delta$ is an ESS if for every
strategy $y \neq x $ there exists some $\bar{\epsilon}_y \in
(0,1)$ such that $u[x,\epsilon y+(1-\epsilon)x]>u[y,\epsilon
y+(1-\epsilon)x]$ holds for all $\epsilon \in (0,
\bar{\epsilon}_y)$ where $\epsilon$ is the proportion of mutant
strategy.
\\
Basically, ESS is a subset of Nash Equilibrium. We use Maynard's
criterion to test whether a Nash Equilibrium is an ESS:\\
$\textbf{Theorem 1. }\Delta^{ESS}=\{x \in
\Delta^{NE}:u(y,y)<y(x,x), \forall y \in \beta^{*}(x), y \neq
x\}.$
\\
To perform all these analysis, we must first define the payoff
matrix of our original game. We will use DP to determine the
numerical payoff values for the four strategy combinations. Assume
each player has a total of \emph{\it{N}} food sources for the
entire development period and in each stage at least \emph{\it{1}}
resource should be consumed in order to maintain basal metabolism.
As we have discussed before, the development period is divided
into three stages: growth initialization, quasilinear growth and
growth termination, hence the player could consume ${1 \cdots
N-2}$ resources in each stage. While the growth is logistic and
nonlinear, we could use linear approximation in each stage as
follows where y is the payoff in each stage and x is the resource
consumed:
\begin{equation}
y=\begin{cases} ax, & \text{Growth Initialization},\\
bx, & \text{Quasilinear Growth},\\
cx, & \text{Growth Termination}.
\end{cases}
\end{equation}
Because logistic curve has a sigmoid shape and is usually
symmetric, it is reasonable to set $a=c$ to reduce computational
intensity. The coefficients $a$ and $b$ has biological meaning of
the efficiency of converting food sources into its own energy and
in our model $b>a$. The DP model is written as follows:
\begin{center}
 $$\text{Maximize }z=\sum^{3}_{i=1}r_ix_i$$
 $$\text{Subject to } \sum^{3}_{i=1}x_i=N$$
\end{center}
The backward DP Formulation for this model is:\\
OVF: $f_k(x)=$ optimal return for the allocation of x units of
resource to stage k $\cdots$ 3.\\
ARG: $(k,x) =$ (stage, units of resource consumed).\\
OPF: $P_k(x)=$ units of resource consumed at stage k.\\
RR:  $f_k(x)=max_{x_k=0,1,\cdots x}(r_kx_k+f_{r+1}(x-x_k))$, $x=1 \cdots N-2$\\
BC:  $f_N(x)=r_N(x)$\\
ANS: $f_1(x)$\\
 For the non-aggressive and non-aggressive strategy
combination, we assume both players do not interfere each other.
In this case, we would only solve the DP for one of them and by
symmetry, the other player should adopt same strategy to maximize
its total payoff. The cost in each stage and state is shown in
Table 2 and we could
calculate the optimal value using DP.\\
For the non-aggressive and aggressive strategy combination, the
cost table is similar to Table 2. The difference is we must define
different $a$ and $b$ values for both strategies. Same thing
happens for the aggressive and aggressive combination. Once we
have figured out the payoff values for each combination we could
complete the payoff matrix and further investigate the ESS.
\begin{center}
\begin{tabular}{c c c c }
  \hline
  State/Stage & 1 & 2 & 3 \\
  \hline
  1 & a & b & a \\
  2 & 2a & 2b & a \\
  $\cdots$ & $\cdots$ & $\cdots$ & $\cdots$ \\
  N-2 & (N-2)a & (N-2)b & (N-2)a \\
  \hline
\end{tabular}
\\
Table 2. Cost Table in Different Stages and States
\end{center}

\section{Model Results}
\subsection{Type I Model: Final Battle} In this simplest case, we
assume both aggressive player and non-aggressive player only fight
after they have depleted all their resources. In real ecosystem,
some animals don't fight while they are young. In fact, they may
even help each other (Taborsky M. 2001)! They fight only when they
are sexually matured. So here we don't even have to bother DP. We
assume the optimal payoffs of non-aggressive and non-aggressive
combination is $(1,1)$ and aggressive player could take advantage
of half of the non-aggressive player's payoff but lose 80\% of its
own payoff when it encounters another aggressive player. The
payoff matrix is:
\begin{center}
\begin{tabular}{c c c}
  \hline
  Strategy & Non-aggressive & Aggressive \\
  \hline
  Non-aggressive & (1,1) & (0.5,1.5) \\
  Aggressive & (1.5,0.5) & (0.2,0.2) \\
  \hline
\end{tabular}
\\
Table 3. Payoff Matrix of Type I Model
\end{center}
In this game there are two pure strategy Nash Equilibria:
nonaggressive- aggressive and aggressive- nonaggressive
combinations. There is another mixed strategy Nash Equilibrium
where both player use aggressive strategy with probability
$\dfrac{5}{8}$ and non-aggressive strategy with probability
$\dfrac{3}{8}$. All these three Nash Equilibria give aggressive
strategy 1.5 payoff and that of non-aggressive is 0.5.\\
According to Smith Maynard's criterion, both pure Nash Equilibria
are not pure ESS and the mixed Nash Equilibrium is the only ESS in
this game.

\subsection{Type II Model: Modified Final Battle} In this modified
final battle model, we assume the fight occurs at the end of
growth as well. Suppose $a=1$ and $b=2$ for non-aggressive player
and aggressive strategy player follows a different development
path. Because most of the development happens during the
quasilinear stage, the simplest case is to assign a smaller
\textit{b} coefficient for aggressive player, for instance,
$b=1.5$. To make our model more realistic, we will modify the
model assumption by taking development lag, development saturation
and both cases into account, thus the second development stage of
aggressive player is not linear. Development lag describes very
slow development when consuming less than certain amount of
resources (lower threshold). Development saturation describes no
development when consuming more than certain amount of resources
(upper threshold). Here we present some typical returns with
respect to different resources
allocation in different development conditions. \\
\begin{center}
\begin{tabular}{c c c c c c c c c}
  \hline
             &   Stage 1   &  & &  & Stage 2 &  & & Stage 3   \\
  \hline
  State/Case & Linear &  Linear & Lag 1 & Lag 2 & Sat. 1 & Sat. 2 & Lag + Sat. & Linear \\
  \hline
  1 & 1 & 1.5 & 1 & 1 & 2 & 2 & 1 & 1 \\
  2 & 2 & 3 & 1 & 1 & 4 & 4 & 2 & 2\\
  3 & 3 & 4.5 & 1 & 1 & 6 & 6 & 4 & 3\\
  4 & 4 & 6 & 2 & 1 & 8 & 6 & 8 & 4\\
  5 & 5 & 7.5 & 4 & 2 & 8 & 6 & 8 & 5\\
  6 & 6 & 9 & 8 & 4 & 8 & 6 & 8 & 6\\
  7 & 7 & 10.5 & 8 & 8 & 8 & 6 & 8 & 7\\
  8 & 8 & 12 & 8 & 8 & 8 & 6 & 8 & 8\\
  \hline
\end{tabular}
\\
Table 4. Return Table in Different Development Conditions for
Aggressive Player
\end{center}

\begin{center}
\begin{tabular}{c c c c c c c c c}
  \hline
             &   Stage 1   & Stage 2 & Stage 3   \\
  \hline
  State/Case & Linear &  Linear & Linear \\
  \hline
  1 & 1 & 2 & 1 \\
  2 & 2 & 4 & 2\\
  3 & 3 & 6 & 3\\
  4 & 4 & 8 & 4\\
  5 & 5 & 10 & 5\\
  6 & 6 & 12 & 6\\
  7 & 7 & 14 & 7\\
  8 & 8 & 16 & 8\\
  \hline
\end{tabular}
\\
Table 5. Return Table for Non-Aggressive Player
\end{center}
Based on table 4, there are 6 different types of return in 2nd
stage for the aggressive player. While for the non-aggressive
player, its returns in different stages are provided in table 5.
So we could formulate a total of 7 DP problems for both players.
Solve these 7 DP problems we have got the optimal allocation
strategy and the corresponding maximum return, shown in table 6.
Please note in table 6 the condition title "linear", "Lag" and
"Saturation" is for the second stage of
entire process only. The notion is different from that of table 4 and 5.\\
\begin{center}
\begin{tabular}{c c c c c c c c}
  \hline
   & Non-aggressive &  &  & Aggressive &  &  \\
  \hline
  Condition & Linear & Linear& Lag 1 & Lag 2 & Sat. 1 & Sat. 2 & Lag + Sat. \\
  \hline
  $P_1(x)$ & 1 & 1 & 1 & 1 & 1 & 1 & 1 \\
  $P_2(x)$ & 8 & 8 & 6 & 7 & 4 & 3 & 4 \\
  $P_3(x)$ & 1 & 1 & 3 & 2 & 5 & 6 & 5 \\
  $f_1(x)$ & 18 & 14 & 12 & 11 & 14 & 13 & 14 \\
  \hline
\end{tabular}
\\
Table 6. Optimal Allocation Strategy and Returns
\end{center}
Since the entire development is symmetric, the resource allocated
in stage 1, $P_1(x)$ and in stage 3, $P_3(x)$ should be
interchangeable. For instance, in Lag 2 condition, the aggressive
player could either consumer 1 unit resource in stage 1 and 3
units of resource in stage 3, or 3 units in stage 1 and 1 unit in
stage 3; the final total returns are just identical.
\\
Now we have got the optimal return so we could calculate the
payoff matrix based on our previous definition. Depending on
different conditions, the total return of aggressive player varies
from 11 to 14. Here we use 12 as instance and calculate the payoff
matrix:\\
\begin{center}
\begin{tabular}{c c c}
  \hline
  Strategy & Non-aggressive & Aggressive \\
  \hline
  Non-aggressive & (18,18) & (9,21) \\
  Aggressive & (21,9) & (2.4,2.4) \\
  \hline
\end{tabular}
\\
Table 7. Payoff Matrix of One Condition of Type II Model
\end{center}
In this game there are three Nash Equilibria, almost the same as
in type I game except that mixed Nash Equilibrium requires
$\dfrac{11}{16}$ non-aggressive and $\dfrac{5}{16}$ aggressive
strategy. From a population point of view, by applying Maynard's
criterion, the mixed Nash Equilibrium is the only ESS in the
evolutionary game.

\subsection{Type III Model: Battles in Every Stage} In this model
the aggressive player will fight in all stages to maximize its
payoff. While it is difficult to model the interaction of fighting
for food source, instead we give the aggressive player larger
coefficients than in Model I and non-aggressive player smaller
coefficients when they encounter. For the aggressive-aggressive
strategy combination, we simply give both players 0 because of
fighting severity. Since they fight in each stage, there is no
final battle in this circumstance. In this model we also consider
two different conditions: $b=0.5$ for non-aggressive player and
$b=1.5$ for aggressive player; $b=1.5$ for non-aggressive player
and $b=2.5$ for aggressive player. The DP results are shown in the
following table:
\begin{center}
\begin{tabular}{c c c c c}
  \hline
   Condition & 1 & 1 & 2 & 2 \\
  \hline
   Player & Non-aggressive & Aggressive & Non-aggressive & Aggressive \\
  \hline
  $P_1(x)$ & 8 & 1 & 1 & 1\\
  $P_2(x)$ & 1 & 8 & 8 & 8\\
  $P_3(x)$ & 1 & 1 & 1 & 1\\
  $f_1(x)$ & 9.5 & 14 & 14 & 22\\
  \hline
\end{tabular}
\\
Table 8. Optimal Allocation Strategy and Returns
\end{center}
So the payoff matrix for condition one is:
\begin{center}
\begin{tabular}{c c c}
  \hline
  Strategy & Non-aggressive & Aggressive \\
  \hline
  Non-aggressive & (18,18) & (9.5,14) \\
  Aggressive & (14,9.5) & (0,0) \\
  \hline
\end{tabular}
\\
Table 9. Payoff Matrix of One Condition of Type III Model
\end{center}
The non-aggressive and non-aggressive strategy combination is the
only Nash Equilibrium in this game; there is no mixed strategy
Nash Equilibrium. This is also the ESS by Maynard's criterion.
\\
 For condition two:
\begin{center}
\begin{tabular}{c c c}
  \hline
  Strategy & Non-aggressive & Aggressive \\
  \hline
  Non-aggressive & (18,18) & (14,22) \\
  Aggressive & (22,14) & (0,0) \\
  \hline
\end{tabular}
\\
Table 10. Payoff Matrix of One Condition of Type III Model
\end{center}
In this game there are three Nash Equilibria, almost the same as
in type I game except that mixed Nash Equilibrium requires
$\dfrac{7}{9}$ non-aggressive and $\dfrac{2}{9}$ aggressive
strategy. From a population point of view, by applying Maynard's
criterion, the mixed Nash Equilibrium is the only ESS in the
evolutionary game.

\section{Discussion}
Though we use DP to find out the optimal allocation strategy, we
have already found under certain circumstances, for instance if
$b>a$, most of our resources should be allocated to the second
stage, the quasilinear growth. In Type II Model we have also
realized the growth does not necessarily be a linear function.
Here we present a criterion to test if we could use the growth
function directly to allocate resources optimally:\\
Assume symmetric growth still holds and define $y=f(x)$ for both
growth initialization and growth termination and $y=g(x)$ for
quasilinear growth. Notice the term "quasilinear growth" here does
not mean the growth function is linear, it could be nonlinear
anyway. If the following is true then we should allocate most of
our resource in the quasilinear growth stage and minimum for
growth initialization and termination: $g(x)$ is not concave and
$g'(x)\geq f'(x), \forall x$.\\
However, this criterion is only sufficient but not necessary. It
is possible to investigate the sufficient and necessary condition
but the computational intensity is almost the same of using DP
because we must compute the first order partial derivative
(gradient) of $f(x)+f(y)+g(10-x-y)$ with respect to $x$ and $y$,
the resource allocated in stage 1 and 3, and determine the
structure of the gradient.\\
In this research project we focus on finite and equal resource
allocation problem for both players. However, our approach could
be extended to unequal resources because we use DP to determine
the optimal strategy for each player so it does not matter whether
the resources are equal for both players. In other words, the
player should not worry about the total amount of their resource
(and actually they cannot determine the amount of resource because
it is pre-specified.) but rather concentrate on how to optimize
the return from the resource (the optimal strategy). It is also
possible to assume infinite resource, however the consumption of
the player is bounded so infinite resource allocation problem
could be transformed to finite resource allocation problem. As we
have discussed before, saturation is a reasonable assumption to
deal with infinite resource. Therefore, we could use DP to solve
almost all types of resources allocation problem for 2 player
game.\\
Another possible improvement of our approach is to introduce
stochastic component into the model. Instead of assigning a
specific amount of resources, we could assume the food resource is
from a certain probability distribution, say, normal distribution.
In effect this is the extension of unequal resource allocation
problem for 2 players. Besides, it is reasonable to assign a
minimum threshold of development and if the player fail to reach
that threshold it then dies. The remaining resources are
transferred to its neighbor (its competitor). In this circumstance
DP could still be applied but we expect the formulation is much
more complicated. When we reach the optimal strategy of resource
allocation we could still apply Game Theory to determine Nash
Equilibrium for a given game but it is difficult to give a close
form representation of what ESS looks like in this scenario
because of stochasticity. We could use simulation to determine the
evolutionary path and this approach is more realistic and useful.

\nocite{Krivan_V} \nocite{Matsumura} \nocite{Nakamaru}
\nocite{Crowley} \nocite{Hardling} \nocite{Al-Tamimi}
\nocite{Hamblin} \nocite{Mesterton} \nocite{Luther} \nocite{Just}
\nocite{Yi} \nocite{Takada} \nocite{Taborsky} \nocite{Neumann}
\nocite{Nash} \nocite{Maynard}

\bibliography{dp}
\bibliographystyle{plain}

\end{document}